\documentclass[12pt]{iopart}
\usepackage{iopams}
\usepackage{graphicx}
\usepackage{dcolumn}
\usepackage{bm}
\usepackage{amssymb}
\usepackage{epstopdf}
\usepackage{color}
\usepackage{soul}

\usepackage[utf8]{inputenc}
\usepackage[T1]{fontenc}
\usepackage{indentfirst}

\usepackage{enumitem}
\usepackage{xspace}

\newcommand{\Td}{$T=\frac12$}
\newcommand{\Tt}{$T=1$}
\newcommand{\abi}{{\it ab initio}}
\newcommand{\thalf}{\tfrac{1}{2}}
\newcommand{\text}{\rm}
\newcommand{\tfrac}[2]{{\textstyle\frac{#1}{#2}}}

\newcommand{\gras}[1]{\boldsymbol{#1}}

\begin{document}

\title{Isobaric Multiplet Mass Equation within nuclear Density Functional Theory}

\author{P. B\k{a}czyk$^1$, W. Satu{\l}a$^{1,2}$, J. Dobaczewski$^{1-4}$ and M. Konieczka$^1$}

\address{$^1$Institute of Theoretical Physics, Faculty of Physics, University of Warsaw, ul. Pasteura 5, PL-02-093 Warsaw, Poland}
\address{$^2$Helsinki Institute of Physics, P.O. Box 64, FI-00014 University of Helsinki, Finland}
\address{$^3$Department of Physics, University of York, Heslington, York YO10 5DD, United Kingdom}
\address{$^4$Department of Physics, PO Box 35 (YFL), FI-40014 University of Jyv{\"a}skyl{\"a}, Finland}

\begin{abstract}
We extend the nuclear Density Functional Theory (DFT) by including
proton-neutron mixing and contact isospin-symmetry-breaking (ISB) terms up to
next-to-leading order (NLO). Within this formalism, we perform
systematic study of the nuclear mirror and triple displacement
energies, or equivalently of the Isobaric Multiplet Mass Equation
(IMME) coefficients. By comparing results with those obtained within
the existing Green Function Monte Carlo (GFMC) calculations, we
address the fundamental question of the physical origin of the
ISB effects. This we achieve by analyzing separate contributions to
IMME coefficients coming from the electromagnetic and nuclear ISB
terms. We show that the ISB DFT and GFMC results agree reasonably
well, and that they describe experimental data with a comparable quality. Since
the separate electromagnetic and nuclear ISB contributions also
agree, we conclude that the beyond-mean-field electromagnetic effects
may not play a dominant role in describing the ISB effects in finite
nuclei.
\end{abstract}


\maketitle


Similarity between the neutron-neutron, proton-proton, and
proton-neutron nuclear interactions was well recognized already in
the third decade of the last century. This property motivated
Heisenberg~\cite{(Hei32b)} and Wigner~\cite{(Wig37a)} to introduce
the concept of isospin symmetry, which abandons notions of protons
and neutrons and replaces them by that of a nucleon, that is, a
particle having two independent states in an abstract space called
isotopic-spin space or, in short, the isospace.

The isospin symmetry is not an exact symmetry of nature.
At the fundamental level, it is violated by the difference in
masses of the constituent $up$ and $down$ quarks and the difference
in their electric charges. At the many-body level, where nucleons are
treated as structureless point-like particles interacting via the
effective forces, the major source of the isospin symmetry breaking
(ISB) is the Coulomb field. The strong-force ISB
components are much weaker than the symmetry conserving, isoscalar ones.
Nevertheless, they are firmly established from the two-body
scattering data, which indicate that the neutron-neutron interaction is
$\sim$1\% stronger than the proton-proton one, and that the
neutron-proton interaction is $\sim$2.5\% stronger than the average of
the former two~\cite{(Mac01a)}.

Following the classification introduced by Henley and
Miller~\cite{(Hen79),(Mil95)}, components of the nuclear force can be
divided into four classes that have different structures with respect
to the isospin symmetry. Apart from the dominant class-I
isoscalar (isospin-invariant) forces, the classification introduces
three different classes of the ISB forces, namely, class-II
isotensor forces, which break the isospin symmetry but are invariant
under a rotation by $\pi$ with respect to the $y-$axis in the
isospace; class-III isovector forces that
break the isospin symmetry but are symmetric under interchange of
nucleonic indices in the isospace, and class-IV forces, which break the isospin
symmetry and, in addition, they mix the total isospin. This
classification is commonly used in the framework of potential models
based on boson-exchange formalism, like CD-Bonn~\cite{(Mac01a)} or
AV18~\cite{(Wir95a),(Wir13)}. It is also a convenient point of
reference for the effective field theory~\cite{(Wal01),(Epe09a),(Orm17)}.

The isospin symmetry is widely used in theoretical modelling of
atomic nuclei. The reason is that the isospin impurity, a measure of
the ISB effect in nuclear wave function, is small -- in heavy $N=Z$
systems of the order of a few percent~\cite{(Sat09a)}. Hence, the
isotopic-spin quantum number $T$ is almost perfectly conserved, and
thus it can be used to classify nuclear many-body states and to work
out selection rules for nuclear reactions.

Although they stem from
small components of nuclear wave functions, the ISB effects manifest
themselves very clearly in the binding energies ($BE<0$) of isobaric
multiplets. This can be visualized by analyzing the mirror (MDE) and
triplet (TDE) displacement energies:
$\mathrm{MDE}=BE\left(A,T,T_z=-T\right)-BE\left(A,T,T_z=+T\right)$ and
$\mathrm{TDE}=BE\left(A,T,T_z=-1\right)+BE\left(A,T,T_z=+1\right)
-2BE\left(A,T,T_z=0\right)$, respectively, where $A$ is the mass number, and
$T$ and $T_z=\frac12(N-Z)$ are the total isospin and its $z$ component.
The MDE and TDE binding-energy indicators were subject to
intensive studies within the shell-model approaches, see~Refs.~\cite{(Orm89),(Zuk02),(Kan17)}
and references quoted therein. This has led to a shell-model identification and
quantification of the effects related to the ISB class-III and II
forces, respectively, for different valence spaces. The class-III terms
have also been studied within the nuclear Density Functional Theory (DFT)
approach~\cite{(Suz93),(Bro00b)}.

In our recent work~\cite{(Bac17b)}, we introduced contact class-II
and III terms simultaneously. Within this approach, which we call ISB
DFT, to treat class-II forces one has to employ the
proton-neutron mixing~\cite{(Sat13d),(She14)}. This allows for
controlling the isospin degree of freedom by the isocranking method,
which can be considered as an approximated isospin
projection~\cite{(Sat13d)}. To determine TDE, such a projection is
indispensable, because in the conventional proton-neutron-unmixed
DFT, states $|T=0,T_z=0\rangle$ and $|T=1,T_z=0\rangle$ are mixed,
whereas only the latter one defines the TDE. The energy of the
unmixed state $|T=1,T_z=0\rangle$ is obtained by
isorotating state $|T=1,T_z=\pm1\rangle$ that
represents isospin-aligned valence particles. The ISB DFT approach allowed us to
reproduce all experimental values of MDEs and TDEs for $A\geq10$.

The goal of
the present Letter is twofold: First, we extend the formalism of
Ref.~\cite{(Bac17b)} from the leading-order (LO) zero-range class~II
and~III interactions to the analogous next-to-leading-order (NLO)
terms. We show that the agreement with data improves, correcting for
the deficiencies of the LO ISB DFT approach identified in~\cite{(Bac17b)}.
Second, we compare our DFT results with those
obtained~\cite{(Wir00a),(Car15)} using an {\abi} Green Function Monte
Carlo (GFMC) approach. This allows us to draw important conclusions
about the role of different components in the ISB sector of
interactions that define properties of finite nuclei.

The NLO (gradient) ISB DFT terms read,
\begin{eqnarray}
\hat{V}_1^{\rm{II}}(i,j) & =& \frac12 t_1^{\rm{II}}
\left( \delta\left( \gras{r}_{ij} \right) \bm{k}^2 + \bm{k}'^2 \delta\left(\gras{r}_{ij} \right) \right)  \hat T^{(ij)}  ,\\
\hat{V}_2^{\rm{II}}(i,j) & = & t_2^{\rm{II}}
\bm{k}' \delta\left(\gras{r}_{ij} \right) \bm{k} \hat T^{(ij)}  , \\
\hat{V}_1^{\rm{III}}(i,j) & =& \frac12 t_1^{\rm{III}}
\left( \delta\left( \gras{r}_{ij} \right) \bm{k}^2 + \bm{k}'^2 \delta\left(\gras{r}_{ij} \right) \right) \hat T_z^{(ij)},\\
\hat{V}_2^{\rm{III}}(i,j) & = & t_2^{\rm{III}}
\bm{k}' \delta\left(\gras{r}_{ij} \right) \bm{k} \hat T_z^{(ij)}  ,
\label{eq:classIII_grad}
\end{eqnarray}
where $\gras{r}_{ij} = \gras{r}_i - \gras{r}_j$,
$\bm{k}  =  \frac{1}{2i}\left(\bm{\nabla}_i-\bm{\nabla}_j\right)$ and
$\bm{k}' = -\frac{1}{2i}\left(\bm{\nabla}_i-\bm{\nabla}_j\right)$ are the standard relative-momentum
operators acting to the right and left, respectively, and
$\hat T^{(ij)} = 3\hat{\tau}_3^{(i)}\hat{\tau}_3^{(j)}-\hat{\vec{\tau}}^{(i)}\circ\hat{\vec{\tau}}^{(j)}$ and
$\hat T_z^{(ij)} = \hat{\tau}_3^{(i)}+\hat{\tau}_3^{(j)}$ are the isotensor and isovector operators.
Similarly as at LO~\cite{(Bac17b)}, the spin-exchange terms are redundant and could be omitted.
The NLO extension brings to the formalism four additional adjustable low-energy coupling constants
(LECs): $t_1^{\rm{II}}$, $t_2^{\rm{II}}$, $t_1^{\rm{III}}$, and $t_2^{\rm{III}}$.

The NLO terms were implemented in the code {\sc
hfodd} (v2.85r)~\cite{(Sch17),(Dob18)}. First, we readjusted the LO LECs of
Ref.~\cite{(Bac17b)} to the available data on MDEs and TDEs in a
wider range of the $A\geq 6$ isospin doublets and
triplets~\cite{(Wan17),(ensdf_url2)}. We also included in the fit the recently
measured mass of $^{44}$V~\cite{(Zha18)}. Similarly as before, we excluded
from the fit several outliers: (i) the $A=9$ and $A=16$ points, which
depend on the $T_z=-T$ masses corresponding to negative proton
separation energies and (ii) the $A=69$ and
$A=73$ points, which depend on masses derived from
systematics~\cite{(Wan17)}.
This defined our dataset used for fitting,
which finally consisted of 32 MDEs for isospin doublets, and 26 MDEs
and 26 TDEs for isospin triplets.

We performed the adjustments using the methodology outlined in detail in
Refs.~\cite{(Dob14b),(Bac17b)}.\footnote{In this Letter, we
corrected a numerical
error of estimating statistical uncertainties that was present
in~Ref.~\cite{(Bac17b)}.} We reiterate
that the fitting of class II and
class III parameters could be done independetly to TDEs and MDEs, respectively.
We used the SV$_{\rm T}$ isospin-conserving
Skyrme functional~\cite{(Bei75b),(Sat14g)}, which is free from unwanted
self-interaction contributions~\cite{(Tar14a)}. Such a fit gave
$t_0^{\rm{II}}=3.7$$\pm$0.4 and
$t_0^{\rm{III}}=-7.3$$\pm$0.3\,MeV\,fm$^3$. This ISB functional is
dubbed  SV$_{\rm T;\, LO}^{\text{ISB}}$. Small differences with
respect to the values obtained in Ref.~\cite{(Bac17b)} are due to a
slightly different dataset used now.

\begin{table}[htb]
\centering
\caption{The root-mean-square deviations (RMSDs) between the DFT and experimental
values of MDEs and TDEs (in keV).}\label{tab:errors}
\begin{tabular}{l|rrr}
    & \multicolumn{1}{c}{~~~~no ISB    }
    & \multicolumn{1}{c}{~~~~ISB at LO }
    & \multicolumn{1}{c}{~~~~ISB at NLO}  \\ \hline
MDE $T=\frac12$ &  547    ~~& 152  ~~~~& 111  ~~~~\\
MDE $T=1$       & 1035    ~~& 330  ~~~~& 180  ~~~~\\
TDE             &  166    ~~&  94  ~~~~&  65  ~~~~\\ \hline
\end{tabular}\label{err-NLO}
\end{table}

\begin{figure}[tb]
\includegraphics[width=\columnwidth]{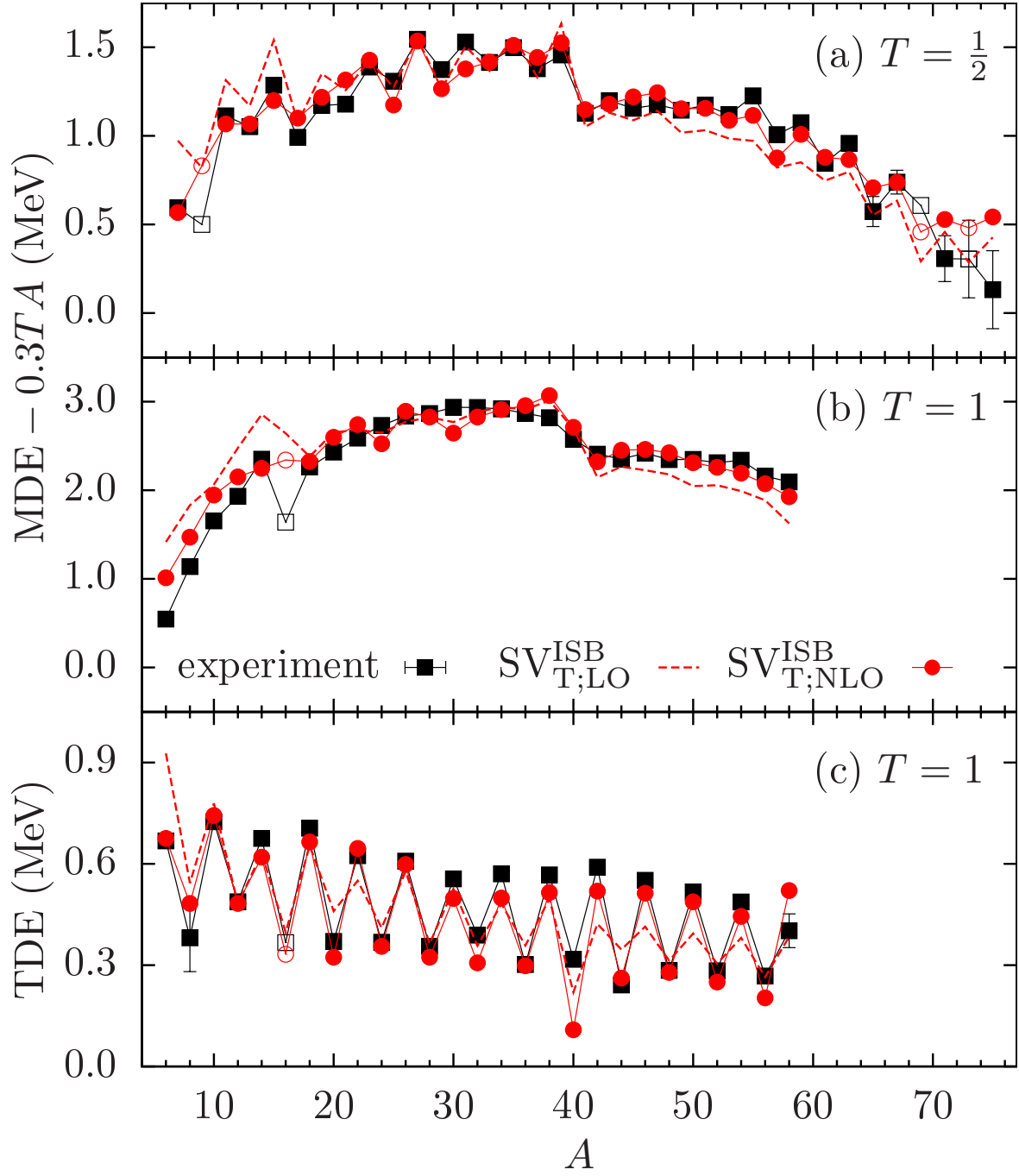}
\caption{(Color online) MDEs in the {\Td} (a) and {\Tt} (b) mirror nuclei, and TDEs (c) in isospin triplets,
calculated using the SV$^{\rm ISB}_{\rm T;LO}$ (dashed line)
and SV$^{\rm ISB}_{\rm T;NLO}$ (circles) functionals, in comparison with experimental data
(squares)~\cite{(Wan17),(ensdf_url2)}. Open symbols denote outliers excluded from the fit, see text.}
\label{fig:MDE&TDE}
\end{figure}

Next, we adjusted all six (LO plus NLO) ISB DFT LECs to the same dataset,
which gave us the ISB functional SV$_{\rm T;\, NLO}^{\text{ISB}}$
defined by the values:
$t_0^{\rm{II}} = -16$$\pm3$ and $t_0^{\rm{III}}= 11$$\pm2$\,MeV\,fm$^3$, and
$t_1^{\rm{II}} =  22$$\pm3$,    $t_1^{\rm{III}}= -14$$\pm4$,
$t_2^{\rm{II}}  =  1$$\pm1$, and $t_2^{\rm{III}} = -7.8$$\pm0.8$\,MeV\,fm$^5$.
The root-mean-square deviations (RMSDs) between calculated and experimental
values of MDEs and TDEs are collected in Table~\ref{err-NLO}, and all
obtained results are plotted in Fig.~\ref{fig:MDE&TDE}.

The NLO terms clearly improve the agreement with data. As hinted on
in~\cite{(Bac17b)}, their surface character allows for better
reproduction of the mass dependence of both MDEs and TDEs. We see
that every next order brings about a factor of 2 of improvement,
which is a characteristic feature of a converging effective
theory~\cite{(Lep97a)}. We also note that values of the LO LECs
adjusted at NLO differ very much from those adjusted at LO. This is
also fully consistent with the rules of the effective theory, and
points to the fact that specific values of the LECs are order
dependent and thus do not carry too much of a physical information.

To find out whether the improvement of the RMSD,
Table~\ref{tab:errors}, is not only a result of the increased number
of adjustable parameters (from 1 to 3 parameters for
each class), we determined the Bayesian Information Criterion (BIC)~\cite{(Pri81)},
\begin{equation}
\mathrm{BIC} = 2n \cdot \ln(\mathrm{RMSD/keV}) + p \cdot \ln(n),
\end{equation}
where $n$ is the number of data points used for fitting and $p$ is
the number of adjustable parameters. This formula for calculating BIC is
valid under the same assumptions that are commonly made when defining the
least-squared fitting approach (independent distribution of the model
errors according to the normal distribution and maximization of the
log likelihood with respect to the true variance). Then, the
difference of BICs between the LO and NLO models reads:
\begin{equation}
\mathrm{BIC_{LO}} - \mathrm{BIC_{NLO}} = 2n \cdot \ln\left(\frac{\mathrm{RMSD_{LO}}}{\mathrm{RMSD_{NLO}}}\right) - 2 \ln(n).
\end{equation}
As it turns out, for class III and II this difference equals 53 and
13, respectively, which strongly favours the NLO model, as it is the
one with much lower value of BIC~\cite{(Kas95)}. To sum up, by
increasing the number of parameters when going from LO to NLO, our
approach leads to a more accurate model that describes the physics
better, rather than to a model that overfits the data.

We can now address the question of what is the nature of
the introduced ISB terms of the functional, namely, whether they model the
strong-force-rooted effects, Coulomb correlations beyond mean field, or
both. In this Letter, we address this question by performing a
systematic study of the Isobaric Multiplet Mass Equation
(IMME)~\cite{(Wig58),(Wei59)}, $BE(A,T,T_z) = a + b T_z + c T_z^2$,
and by comparing the DFT results with those obtained within the GFMC approach~\cite{(Wir00a),(Car15)}.

The quadratic dependence
of binding energies on $T_z$, which is assumed in IMME, is motivated by the
expansion of the two-body Coulomb force
into isoscalar, isovector, and isotensor terms~\cite{(Pes61a)}, which
also motivates the following IMME variant~\footnote{The
minus sign in the second term conforms with
the opposite sign in the definition of $T_z$ used in Refs.~\cite{(Wir00a),(Car15)}.},
\begin{eqnarray}
BE(A,T,I,T_z ) &=& a^{(0)}_{A,T,I}-a^{(1)}_{A,T,I}T_z + \thalf a^{(2)}_{A,T,I}\{ 3 T_z^2 - T(T+1) \},
\label{IMME3}
\end{eqnarray}
where for the IMME coefficients $a^{(n)}_{A,T,I}$ we used the
traditional notation that includes the angular-momentum quantum
number $I$.

Of course, for triplets ($T=1$), nuclear masses can always be
trivially described by a parabolic dependence on $T_z$, so then the
information carried by the IMME coefficients is exactly the same as
that contained in the displacement energies discussed above, with
$\mathrm{MDE}=2Ta^{(1)}_{A,T,I}=-2Tb$ and $\mathrm{TDE}=
3T^2a^{(2)}_{A,T,I}=2T^2c$. For higher multiplets ($T>1$), there is an
active ongoing debate if higher-order terms, proportional to $T_z^3$
or $T_z^4$, are required, see Refs.~\cite{(Nes17b),(Bro17)} for brief
recent reviews.

We investigate the intrinsic structure of the IMME coefficients by decomposing them into
contributions coming from the electromagnetic and contact ISB
parts of the functional. We compare the DFT results with those
obtained using the GFMC method, split into the electromagnetic and nuclear ISB parts, see
Refs.~\cite{(Wir00a),(Wir13),(Car15)}. In this way, we try to bridge the gap between our
phenomenological ISB terms with LECs fitted to many-body data and the AV18~\cite{(Wir95a)} ISB
forces with LECs adjusted to two-body scattering data. The GFMC calculations involve
high-precision potential AV18, which takes into account
the ISB effects due to the one-photon and higher-order
electromagnetic effects, isovector kinetic energy, and class-II and class-III strong
finite-range regularized interactions. On the other hand, our DFT modelling
captures all ISB effects (beyond the mean-field Coulomb) either in two (LO) or six
(NLO) LECs corresponding to contact class-II and class-III forces.

\begin{table}[htb]
\centering
\caption{Contributions of the electromagnetic ($V^\gamma$), ISB
nuclear ($V^{\rm{ISB}}$), and isoscalar ($H^{\rm{T=0}}$) forces to
coefficients $a^{(1)}_{A,T,I}$, see Eq.~(\protect\ref{IMME3}), calculated using the
GFMC~\protect\cite{(Car15)} and ISB DFT  at LO and NLO.
Theoretical uncertainties of the DFT results are related to the uncertainties of the
adjusted LECs. All values are in keV.}
\label{tab:IMME-1}
\begin{tabular}{lcr@{}lr@{}lr@{}lr@{}lr}
\hline
$a^{{(1)}}_{A,T,I}$ & Model       & \multicolumn{2}{c}{$V^\gamma$}
                                  & \multicolumn{2}{c}{$V^{\rm{ISB}}$}
                                  & \multicolumn{2}{c}{$H^{\rm{T=0}}$}
                                  & \multicolumn{2}{c}{Total}     & EXP   \\
\hline
                    &GFMC                        &   1056&(1)  &   62&(0)  &   68&(3)  &   1184&(4)     &       \\
$a^{(1)}_{6,1,0}$   &SV$_{\rm{T;LO}}^{\text{ISB}}$     &    1278&     &   320&     &   11&     &   1609&(15)    &   1174   \\
                    &SV$_{\rm{T;NLO}}^{\text{ISB}}$     &    1277&     &   118&     &   12&     &   1407&(24)    &       \\
\hline
                    &GFMC                        &   1478&(2)  &   106&(1)  &   27&(10)  &   1611&(10)     &       \\
$a^{(1)}_{7,\frac12,\frac32}$   &SV$_{\rm{T;LO}}^{\text{ISB}}$     &    1539&     &   472&     &   13&     &   2024&(25)    &   1644    \\
                    &SV$_{\rm{T;NLO}}^{\text{ISB}}$     &    1537&     &   66&     &   14&     &   1617&(48)    &       \\
\hline
                    &GFMC                        &    1675&(1)  &   102&(1)  &   43&(6)  &   1813&(6)     &       \\
$a^{(1)}_{8,1,2}$   &SV$_{\rm{T;LO}}^{\text{ISB}}$     &    1730&     &   368&     &   18&     &   2116&(17)    &   1770    \\
                    &SV$_{\rm{T;NLO}}^{\text{ISB}}$     &    1726&     &   189&     &   21&     &   1936&(21)    &       \\
\hline
                    &GFMC                        &    2155&(7)  &   110&(1)  &  ---&     &   2170&(8)     &       \\
$a^{(1)}_{10,1,0}$  &SV$_{\rm{T;LO}}^{\text{ISB}}$  &    2154&     &   354&     &   25&     &   2533&(16)    & 2329  \\
                    &SV$_{\rm{T;NLO}}^{\text{ISB}}$  &    2146&     &   296&     &   31&     &   2474&(13)    &       \\
\hline
                    &  SV$_{\rm{T;LO}}^{\text{ISB}}$&    2432&  &  505&  &   26&  &    2964&(26)     &       \\[-1.0ex]
$a^{(1)}_{11,\frac12,\frac32}$   &                            &       &     &      &     &     &     &       &        &  2764  \\[-1.4ex]
                    &  SV$_{\rm{T;NLO}}^{\text{ISB}}$&     2424&     &    260&     &    33&     &    2717&(33)    &       \\[0.6ex]
\hline
                    &  SV$_{\rm{T;LO}}^{\text{ISB}}$&    2589&  &  419&  &   31&  &    3040&(19)     &       \\[-1.0ex]
$a^{(1)}_{12,1,1}$   &                            &       &     &      &     &     &     &       &        &  2767  \\[-1.4ex]
                    &  SV$_{\rm{T;NLO}}^{\text{ISB}}$&     2584&     &    257&     &    35&     &    2876&(21)    &       \\[0.6ex]
\hline
                    &  SV$_{\rm{T;LO}}^{\text{ISB}}$&    2736&  &  345&  &   39&  &    3120&(21)     &       \\[-1.0ex]
$a^{(1)}_{13,\frac12,\frac12}$   &                            &       &     &      &     &     &     &       &        &  3003  \\[-1.4ex]
                    &  SV$_{\rm{T;NLO}}^{\text{ISB}}$&     2736&     &    244&     &    38&     &    3018&(33)    &       \\[0.6ex]
\hline
                    &  SV$_{\rm{T;LO}}^{\text{ISB}}$&    3006&  &  489&  &   34&  &    3529&(22)     &       \\[-1.2ex]
$a^{(1)}_{14,1,0}$   &                            &       &     &      &     &     &     &       &        &  3276  \\[-0.6ex]
                    &  SV$_{\rm{T;NLO}}^{\text{ISB}}$&     3002&     &    185&     &    38&     &    3225&(34)    &       \\
\hline
\end{tabular}
\end{table}

\begin{table}[htb]
\centering
\caption{Same as in Table~\protect\ref{tab:IMME-1} but for
coefficients $a^{(2)}_{A,T,I}$.}
\label{tab:IMME-2}
\begin{tabular}{lcr@{}lr@{}lr@{}lr@{}lr}
\hline
$a^{{(2)}}_{A,T,I}$ &Model        & \multicolumn{2}{c}{$V^\gamma$}
                                  & \multicolumn{2}{c}{$V^{\rm{ISB}}$}
                                  & \multicolumn{2}{c}{$H^{\rm{T=0}}$}
                                  & \multicolumn{2}{c}{Total}     & EXP   \\
\hline
                    &  GFMC                      &    153&(1)  &  112&(2)  &   5&(4)  &    270&(5)     &       \\
$a^{(2)}_{6,1,0}$   &  SV$_{\rm{T;LO}}^{\text{ISB}}$    &     171&     &    132&     &    7&     &    309&(17)    &   223    \\
                    &  SV$_{\rm{T;NLO}}^{\text{ISB}}$&     165&     &      43&     &  17&     &    225&(16)    &       \\
\hline
                    &  GFMC                      &    136&(1)  &  $-$3&(2)  &   10&(5)  &    139&(5)     &       \\
$a^{(2)}_{8,1,2}$   &  SV$_{\rm{T;LO}}^{\text{ISB}}$&     146&     &    27&     &    8&     &    181&(12)    &   127    \\
                    &  SV$_{\rm{T;NLO}}^{\text{ISB}}$&     142&     &    5&     &    15&     &    161&(18)    &       \\
\hline
                    &  GFMC                      &    178&(1)  &   119&(18) &  ---&     &    297&(19)    &       \\
$a^{(2)}_{10,1,0}$  &  SV$_{\rm{T;LO}}^{\text{ISB}}$&     156&     &    94&     &   10&     &    260&(13)    &   242    \\
                    &  SV$_{\rm{T;NLO}}^{\text{ISB}}$&     146&     &    76&     &   26&     &    248&(9)    &       \\
\hline
$a^{(2)}_{12,1,1}$  &  SV$_{\rm{T;LO}}^{\text{ISB}}$&    135&     &    19&     &    7&     &    160&(6)    &  162  \\
                                &  SV$_{\rm{T;NLO}}^{\text{ISB}}$&    134&     &    15&     &   12&     &    161&(10)    &    \\
\hline
$a^{(2)}_{14,1,0}$  &  SV$_{\rm{T;LO}}^{\text{ISB}}$&    146&     &    68&     &   -4&     &    211&(10)    &  225  \\
                                &  SV$_{\rm{T;NLO}}^{\text{ISB}}$&    142&     &    67&     &   -3&     &    207&(7)    &    \\
\hline
\end{tabular}
\end{table}

\begin{figure}[tb]
\includegraphics[width=\columnwidth]{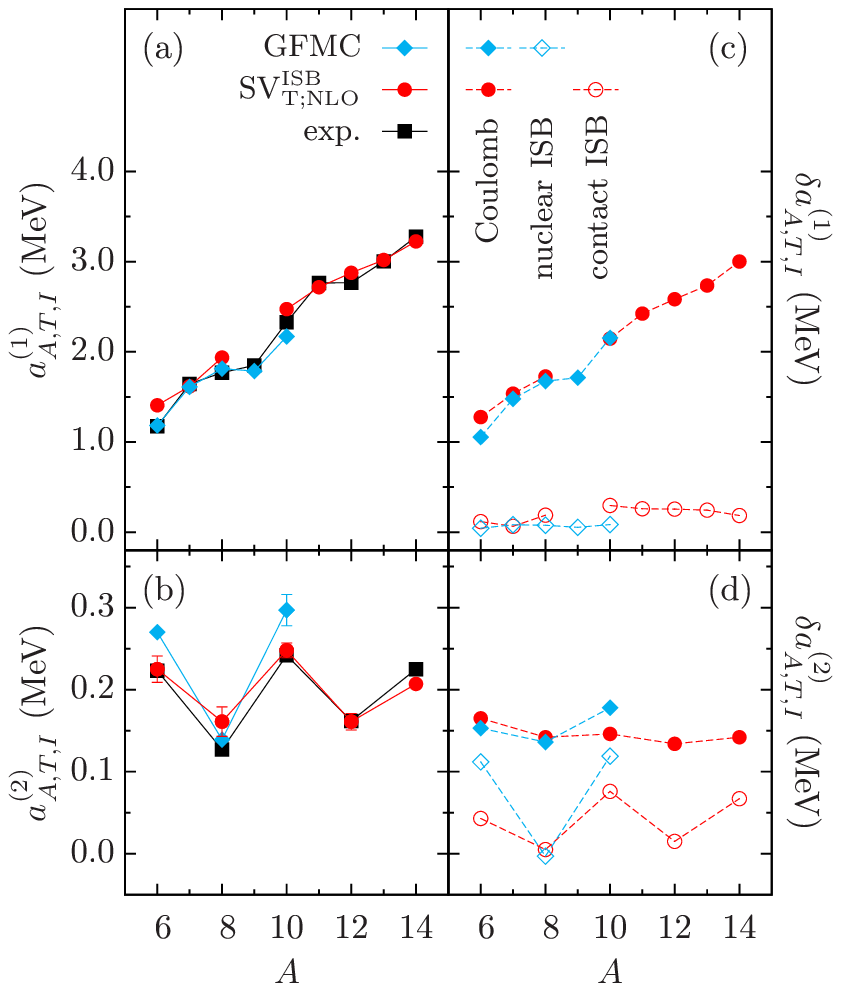}
\caption{(Color online) IMME coefficients $a^{(1)}_{A,T,I}$
(a) and $a^{(2)}_{A,T,I}$ (b), see Eq.~(\protect\ref{IMME3}),
and contributions
to the IMME coefficients ${\delta}a^{(1)}_{A,T,I}$ (c) and
${\delta}a^{(2)}_{A,T,I}$ (d) due to electromagnetic (full symbols),
nuclear ISB (open diamonds), and contact ISB (open circles) forces. For every $A$, quantum numbers
$T$ and $I$ are listed in Tables~\protect\ref{tab:IMME-1} and~\protect\ref{tab:IMME-2}.}
\label{fig:a1a2}
\end{figure}

The DFT and GFMC results are collected in Tables~\ref{tab:IMME-1}
and~\ref{tab:IMME-2} together with experimental
data~\cite{(Wan17),(ensdf_url2)}. As shown in the Tables and
visualized in Figs.~\ref{fig:a1a2}(a) and~\ref{fig:a1a2}(b), the DFT
and GFMC results are of a comparable quality. In the two lightest
triplets ($A=6$ and 8), coefficients $a^{(1)}_{A,T,I}$ are
slightly better described by the GFMC. This approach, unlike the DFT,
probably takes better into account the continuum effects that may be
present here due the proximity of the proton-emission threshold.
Indeed, in $^6$Be (S$_p$=590keV) and  $^8$B (S$_p$=136keV),
relatively large Thomas-Ehrman
shifts~\cite{(Tho51a),(Ehr51a),(Tho52a)} are seen in the
spectra~\cite{(ensdf_url2)}.

In the same two lightest triplets, coefficients $a^{(2)}_{A,T,I}$ are
better described by the ISB DFT. It should be said,
however, that the experimental value of $a^{(2)}_{8,1,2}$ is somewhat uncertain, as
it relays on a model-dependent evaluation of the isospin mixing in  a
near-degenerate doublet of $I=2^+$ states at 16.626 and 16.922\,MeV
in $^8$Be. In this work, following the analysis of
Ref.~\cite{(Car15)}, we adopted the value of 16.8\,MeV for the
excitation energy of the so called {\it empirical\/} $T$=1 state and,
consequently, 128\,keV for the value of $a^{(2)}_{8,1,2}$. Taking instead the
value of 16.724\,MeV, which results from a simple two-level mixing model,
would lead to $a^{(2)}_{8,1,2}$=0.179\,keV.

It is striking that, except for $A$=6, both models predict almost the same
contributions to $a^{(1)}_{A,T,I}$ coming from the electromagnetic
force, see Fig.~\ref{fig:a1a2}(c). Since the GFMC approach contains Coulomb
correlations beyond mean field and DFT does not, this may hint to the fact
that such correlations may not be essential in reproducing the ISB effects
in the many-body context.

Except for the $A$=10 triplet, the nuclear ISB contributions in GFMC are also
consistent with contact ISB contributions in DFT. Note, however, that in this case the GFMC (DFT) underestimates
(overestimates) the $a^{(1)}_{10,1,0}$ coefficient by a similar amount. It is likely, that the
differences reflect a too weak (too strong) nuclear (contact) ISB forces in the GFMC (DFT) calculations.
This consistency supports the interpretation that the contact ISB DFT terms in fact describes mostly the nuclear
ISB effects.
Note in addition that at LO, the ISB DFT contributions to the isovector coefficients are almost
three times larger than the GFMC values.  This underlines the importance of the NLO class-III corrections
introduced in this work, which seem to be indispensable for a proper treatment of the ISB effects
in the isovector channel.

The NLO ISB DFT terms are equally important for description of the isotensor
channel. Indeed, the isotensor coefficients $a^{(2)}_{A,T,I}$ are
exceptionally well reproduced at NLO, even better than in
the GFMC calculations, see Fig.~\ref{fig:a1a2}(b).  The individual
contributions due to electromagnetic and nuclear/contact in GFMC/DFT ISB effects, see
Fig.~\ref{fig:a1a2}(d), are comparable in both models.

\begin{figure}[tb]
\includegraphics[width=\columnwidth]{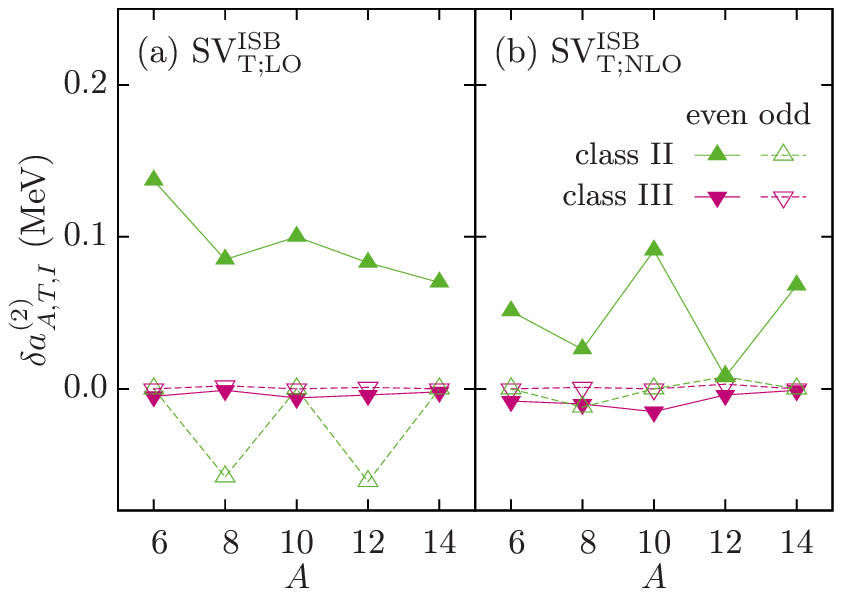}
\caption{(Color online) Contributions to the ISB DFT IMME
coefficients ${\delta}a^{(2)}_{A,T,I}$ due to time-even (full symbols) and time-odd (open
symbols) class-II (up triangles) and class-III (down triangles) mean fields,
determined at LO (a) and NLO (b).}
\label{fig:a2_TE_TO1}
\end{figure}

In the DFT calculations, the staggering in
$a^{(2)}_{A,T,I}$, Fig.~\ref{fig:a1a2}(d), comes entirely from the
contact class-II force. The details depend, however, on the order of
approximation. At LO (NLO), the staggering is due to the
class-II time-odd (time-even) mean fields, see
Figs.~\ref{fig:a2_TE_TO1}. In
the GFMC calculations, both the electromagnetic and strong
class-II forces contribute to the staggering, but the latter
contribution prevails. We also note here that in the nuclear shell model,
the staggering is explained in terms of the $J=0$ neutron-proton
pairing, see Ref.~\cite{(Zuk02)}.

\begin{figure}[tb]
\includegraphics[width=\columnwidth]{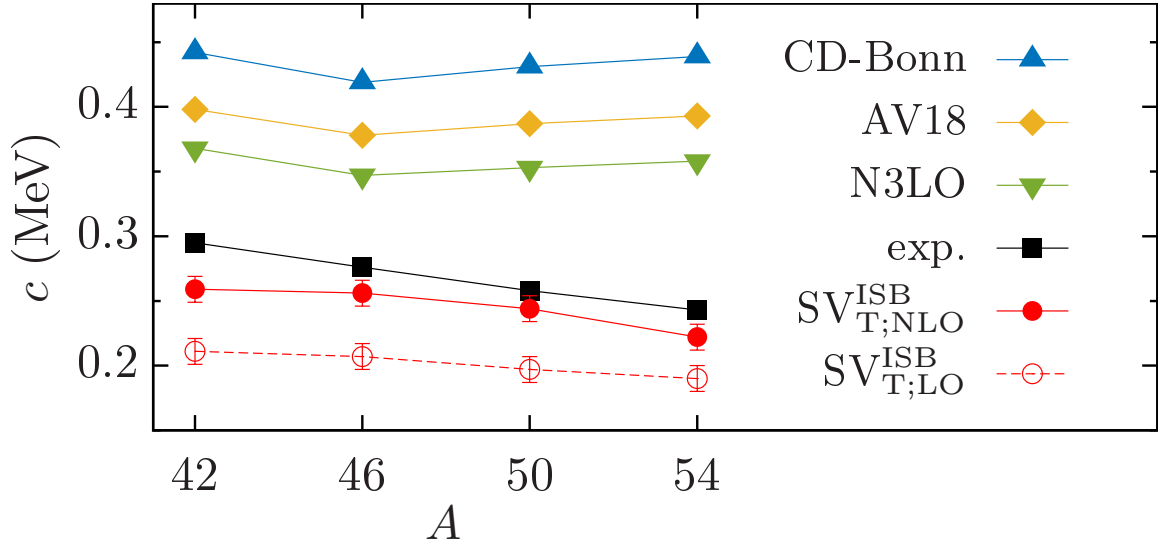}
\caption{(Color online) Isotensor coefficients $c$ calculated within
the ISB DFT and shell-model-based {\abi} approach
at 3$^{\rm rd}$ order~\protect\cite{(Orm17)}.}
\label{fig:a2_TE_TO2}
\end{figure}

Recently, Ormand {\it et al.\/}~\cite{(Orm17)} performed shell-model-based
{\abi} study of the isotensor IMME coefficients $c$ in the
$pf$-shell isospin triplets ($A=42$, 46, 50, and 54). Their results
systematically overestimate the experimental values, irrespective of
which high-precision potential, CD-Bonn~\cite{(Mac01a)},
AV18~\cite{(Wir95a)}, or N3LO~\cite{(Epe09a)}, was used in the
calculations, see Fig.~\ref{fig:a2_TE_TO2}. Conversely, our ISB DFT
calculations at LO and NLO systematically underestimate the
experimental data, but at NLO the level of agreement greatly
improves. The DFT and shell-model-based {\abi} methods predict very
different contributions to $c$ due to the Coulomb and contact/nuclear ISB forces. At
present it is impossible, however, to draw deeper conclusions,
because the results of Ref.~\cite{(Orm17)} do not seem yet to
converge with respect to the order of calculation, and because those
for the $A=4n$ triplets are not yet available. The DFT results
presented in this Letter may thus serve as a baseline for future
comparisons of both approaches in heavy nuclei.

The ISB character of the strong interaction manifests itself not only
in the values of MDE and TDE. One can expect it to be important in
describing $\beta$ decays, differences of energies of levels of
nuclei in isospin multiplets (mirror and triplet energy differences,
MED and TED), giant and Gamow-Teller resonances, etc. The ISB DFT
seems to be a perfect tool to study all these observables. Currently
our group is working on the effects of the short-range ISB terms of
class III on the $\beta$ decays in mirror nuclei and on MEDs in
rotational bands. The results will be published in forthcoming
publications.

In summary, we performed systematic study of mirror and triplet
displacement energies, or equivalently, isovector and isotensor IMME
coefficients, using the extended DFT approach that includes
proton-neutron mixing and contact ISB terms at LO and NLO. In light nuclei,
we compared the obtained IMME coefficients with the results of
existing GFMC calculations. We focused on comparing partial
contributions due to electromagnetic and nuclear ISB terms.

We showed that the NLO terms greatly improve the agreement of the DFT
results with data, and that the DFT and GFMC calculations reproduce
empirical IMME coefficients comparably well. But most importantly, we
showed that the Coulomb contributions to the IMME coefficients are
similar in both approaches, which implies that the Coulomb
correlations beyond mean field may not be crucial in reproducing the ISB effects
in the many-body context.

We thank the authors of Ref.~\cite{(Orm17)} for sending us numerical
values of their results. This work was supported in part by the
Polish National Science Centre under Contract
Nos.~2014/15/N/ST2/03454 and 2015/17/N/ST2/04025, by the Academy of
Finland and University of Jyv\"askyl\"a within the FIDIPRO program,
and by the STFC Grants No.~ST/M006433/1 and No.~ST/P003885/1. We
acknowledge the CSC-IT Center for Science Ltd., Finland, and CI\'S
\'Swierk Computing Center, Poland, for the allocation of
computational resources.


\newpage


\begin{thebibliography}{10}
\expandafter\ifx\csname url\endcsname\relax
  \def\url#1{{\tt #1}}\fi
\expandafter\ifx\csname urlprefix\endcsname\relax\def\urlprefix{URL }\fi
\providecommand{\eprint}[2][]{\url{#2}}

\bibitem{(Hei32b)}
Heisenberg W 1932 {\em Zeitschrift f{\"u}r Physik\/} {\bf 77} 1--11
  \urlprefix\url{https://doi.org/10.1007/BF01342433}

\bibitem{(Wig37a)}
Wigner E 1937 {\em Phys. Rev.\/} {\bf 51}(2) 106
  \urlprefix\url{http://link.aps.org/doi/10.1103/PhysRev.51.106}

\bibitem{(Mac01a)}
Machleidt R 2001 {\em Phys. Rev. C\/} {\bf 63} 024001
  \urlprefix\url{\newline{}https://link.aps.org/doi/10.1103/PhysRevC.63.024001}

\bibitem{(Hen79)}
Henley E~M and Miller G~A 1979 {\em Mesons in Nuclei\/} (North Holland)

\bibitem{(Mil95)}
Miller G~A and van Oers W~H~T 1995 {\em Symmetries and Fundamental Interactions
  in Nuclei\/} (World Scientific)

\bibitem{(Wir95a)}
Wiringa R~B, Stoks V~G~J and Schiavilla R 1995 {\em Phys. Rev. C\/} {\bf 51}(1)
  38--51 \urlprefix\url{https://link.aps.org/doi/10.1103/PhysRevC.51.38}

\bibitem{(Wir13)}
Wiringa R~B, Pastore S, Pieper S~C and Miller G~A 2013 {\em Phys. Rev. C\/}
  {\bf 88}(4) 044333
  \urlprefix\url{http://link.aps.org/doi/10.1103/PhysRevC.88.044333}

\bibitem{(Wal01)}
Walzl M, Meißner U~G and Epelbaum E 2001 {\em Nuclear Physics A\/} {\bf 693}
  663 -- 692 ISSN 0375-9474
  \urlprefix\url{http://www.sciencedirect.com/science/article/pii/S0375947401009691}

\bibitem{(Epe09a)}
Epelbaum E, Hammer H~W and Mei\ss{}ner U~G 2009 {\em Rev. Mod. Phys.\/} {\bf
  81}(4) 1773--1825
  \urlprefix\url{https://link.aps.org/doi/10.1103/RevModPhys.81.1773}

\bibitem{(Orm17)}
Ormand W~E, Brown B~A and Hjorth-Jensen M 2017 {\em Phys. Rev. C\/} {\bf 96}(2)
  024323 \urlprefix\url{https://link.aps.org/doi/10.1103/PhysRevC.96.024323}

\bibitem{(Sat09a)}
Satu\l{}a W, Dobaczewski J, Nazarewicz W and Rafalski M 2009 {\em Phys. Rev.
  Lett.\/} {\bf 103}(1) 012502
  \urlprefix\url{http://link.aps.org/doi/10.1103/PhysRevLett.103.012502}

\bibitem{(Orm89)}
Ormand W and Brown B 1989 {\em Nucl. Phys. A\/} {\bf 491} 1 ISSN 0375-9474
  \urlprefix\url{http://www.sciencedirect.com/science/article/pii/0375947489902030}

\bibitem{(Zuk02)}
Zuker A~P, Lenzi S~M, Martinez-Pinedo G and Poves A 2002 {\em Phys. Rev.
  Lett.\/} {\bf 89}(14) 142502
  \urlprefix\url{https://link.aps.org/doi/10.1103/PhysRevLett.89.142502}

\bibitem{(Kan17)}
Kaneko K, Sun Y, Mizusaki T, Tazaki S and Ghorui S 2017 {\em Physics Letters
  B\/} {\bf 773} 521 -- 526 ISSN 0370-2693
  \urlprefix\url{http://www.sciencedirect.com/science/article/pii/S0370269317306780}

\bibitem{(Suz93)}
Suzuki T, Sagawa H and Van~Giai N 1993 {\em Phys. Rev. C\/} {\bf 47}(4)
  R1360--R1363
  \urlprefix\url{http://link.aps.org/doi/10.1103/PhysRevC.47.R1360}

\bibitem{(Bro00b)}
Brown B~A, Richter W~A and Lindsay R 2000 {\em Physics Letters B\/} {\bf 483}
  49 -- 54 ISSN 0370-2693
  \urlprefix\url{http://www.sciencedirect.com/science/article/pii/S037026930000589X}

\bibitem{(Bac17b)}
B\k{a}czyk P, Dobaczewski J, Konieczka M, Satu\l{}a W, Nakatsukasa T and Sato K
  2018 {\em Physics Letters B\/} {\bf 778} 178 -- 183
  \urlprefix\url{https://arxiv.org/abs/1701.04628v3}

\bibitem{(Sat13d)}
Sato K, Dobaczewski J, Nakatsukasa T and Satu\l{}a W 2013 {\em Phys. Rev. C\/}
  {\bf 88}(6) 061301
  \urlprefix\url{https://link.aps.org/doi/10.1103/PhysRevC.88.061301}

\bibitem{(She14)}
Sheikh J~A, Hinohara N, Dobaczewski J, Nakatsukasa T, Nazarewicz W and Sato K
  2014 {\em Phys. Rev. C\/} {\bf 89}(5) 054317
  \urlprefix\url{http://link.aps.org/doi/10.1103/PhysRevC.89.054317}

\bibitem{(Wir00a)}
Wiringa R~B, Pieper S~C, Carlson J and Pandharipande V~R 2000 {\em Phys. Rev.
  C\/} {\bf 62}(1) 014001
  \urlprefix\url{https://link.aps.org/doi/10.1103/PhysRevC.62.014001}

\bibitem{(Car15)}
Carlson J, Gandolfi S, Pederiva F, Pieper S~C, Schiavilla R, Schmidt K~E and
  Wiringa R~B 2015 {\em Rev. Mod. Phys.\/} {\bf 87}(3) 1067--1118
  \urlprefix\url{http://link.aps.org/doi/10.1103/RevModPhys.87.1067}

\bibitem{(Sch17)}
Schunck N, Dobaczewski J, Satu{\l}a W, B\k{a}czyk P, Dudek J, Gao Y, Konieczka
  M, Sato K, Shi Y, Wang X and Werner T 2017 {\em Computer Physics
  Communications\/} {\bf 216} 145 -- 174 ISSN 0010-4655
  \urlprefix\url{http://www.sciencedirect.com/science/article/pii/S0010465517300942}

\bibitem{(Dob18)}
Dobaczewski J {\it et al.}, to be published

\bibitem{(Wan17)}
Wang M, Audi G, Kondev F, Huang W, Naimi S and Xu X 2017 {\em Chinese Physics
  C\/} {\bf 41} 030003
  \urlprefix\url{http://stacks.iop.org/1674-1137/41/i=3/a=030003}

\bibitem{(ensdf_url2)}
Evaluated Nuclear Structure Data File, http://www.nndc.bnl.gov/ensdf/
  \urlprefix\url{http://www.nndc.bnl.gov/ensdf/}

\bibitem{(Zha18)}
Zhang Y~H, Zhang P, Zhou X~H, Wang M, Litvinov Y~A, Xu H~S, Xu X, Shuai P, Lam
  Y~H, Chen R~J, Yan X~L, Bao T, Chen X~C, Chen H, Fu C~Y, He J~J, Kubono S,
  Liu D~W, Mao R~S, Ma X~W, Sun M~Z, Tu X~L, Xing Y~M, Zeng Q, Zhou X, Zhan
  W~L, Litvinov S, Blaum K, Audi G, Uesaka T, Yamaguchi Y, Yamaguchi T, Ozawa
  A, Sun B~H, Sun Y and Xu F~R 2018 {\em Phys. Rev. C\/} {\bf 98}(1) 014319
  \urlprefix\url{https://link.aps.org/doi/10.1103/PhysRevC.98.014319}

\bibitem{(Dob14b)}
Dobaczewski J, Nazarewicz W and Reinhard P~G 2014 {\em Journal of Physics G:
  Nuclear and Particle Physics\/} {\bf 41} 074001
  \urlprefix\url{http://stacks.iop.org/0954-3899/41/i=7/a=074001}

\bibitem{(Bei75b)}
Beiner M, Flocard H, Giai N~V and Quentin P 1975 {\em Nuclear Physics A\/} {\bf
  238} 29 -- 69 ISSN 0375-9474
  \urlprefix\url{http://www.sciencedirect.com/science/article/pii/0375947475903383}

\bibitem{(Sat14g)}
Satu\l{}a W and Dobaczewski J 2014 {\em Phys. Rev. C\/} {\bf 90}(5) 054303
  \urlprefix\url{https://link.aps.org/doi/10.1103/PhysRevC.90.054303}

\bibitem{(Tar14a)}
Tarpanov D, Toivanen J, Dobaczewski J and Carlsson B~G 2014 {\em Phys. Rev.
  C\/} {\bf 89}(1) 014307
  \urlprefix\url{http://link.aps.org/doi/10.1103/PhysRevC.89.014307}

\bibitem{(Lep97a)}
Lepage G 1997 {\em nucl-th/9706029\/}
  \urlprefix\url{https://arxiv.org/abs/nucl-th/9706029}

\bibitem{(Pri81)}
Priestley M~B 1981 {\em Spectral Analysis and Time Series\/} (Academic Press) p
  375

\bibitem{(Kas95)}
Kass R~E and Raftery A~E 1995 {\em Journal of the American Statistical
  Association\/} {\bf 90} 773--795
  \urlprefix\url{https://www.tandfonline.com/doi/abs/10.1080/01621459.1995.10476572}

\bibitem{(Wig58)}
Wigner E~P 1958 {\em Proceedings of the Robert A. Welsch Conference on Chemical
  Research, Vol. 1\/} ed Milligan W~D (R. A. Welsch Foundation, Houston, TX)
  p~88

\bibitem{(Wei59)}
Weinberg S and Treiman S~B 1959 {\em Phys. Rev.\/} {\bf 116}(2) 465--468
  \urlprefix\url{https://link.aps.org/doi/10.1103/PhysRev.116.465}

\bibitem{(Pes61a)}
Peshkin M 1961 {\em Phys. Rev.\/} {\bf 121}(2) 636
  \urlprefix\url{\newline{}http://link.aps.org/doi/10.1103/PhysRev.121.636}

\bibitem{(Nes17b)}
Nesterenko D~A, Kankainen A, Canete L, Block M, Cox D, Eronen T, Fahlander C,
  Forsberg U, Gerl J, Golubev P, Hakala J, Jokinen A, Kolhinen V~S, Koponen J,
  Lalovi\'c N, Lorenz C, Moore I~D, Papadakis P, Reinikainen J, Rinta-Antila S,
  Rudolph D, Sarmiento L~G, Voss A and \"Ayst\"o J 2017 {\em Journal of Physics
  G: Nuclear and Particle Physics\/} {\bf 44} 065103
  \urlprefix\url{http://stacks.iop.org/0954-3899/44/i=6/a=065103}

\bibitem{(Bro17)}
Brodeur M, Kwiatkowski A~A, Drozdowski O~M, Andreoiu C, Burdette D, Chaudhuri
  A, Chowdhury U, Gallant A~T, Grossheim A, Gwinner G, Heggen H, Holt J~D,
  Klawitter R, Lassen J, Leach K~G, Lennarz A, Nicoloff C, Raeder S, Schultz
  B~E, Stroberg S~R, Teigelh\"ofer A, Thompson R, Wieser M and Dilling J 2017
  {\em Phys. Rev. C\/} {\bf 96}(3) 034316
  \urlprefix\url{https://link.aps.org/doi/10.1103/PhysRevC.96.034316}

\bibitem{(Tho51a)}
Thomas R~G 1951 {\em Phys. Rev.\/} {\bf 81} 148
  \urlprefix\url{\newline{}https://link.aps.org/doi/10.1103/PhysRev.81.148}

\bibitem{(Ehr51a)}
Ehrman J~B 1951 {\em Phys. Rev.\/} {\bf 81} 412
  \urlprefix\url{http://link.aps.org/doi/10.1103/PhysRev.81.412}

\bibitem{(Tho52a)}
Thomas R~G 1952 {\em Phys. Rev.\/} {\bf 88} 1109
  \urlprefix\url{\newline{}http://link.aps.org/doi/10.1103/PhysRev.88.1109}

\end{thebibliography}

\providecommand{\newblock}{}

\end{document}